\documentclass{statement}

\usepackage{tcolorbox}

\usepackage{palatino}  
\usepackage[T1]{fontenc}
\definecolor{linkcol}{rgb}{0.278,0.541,0.459}%

\definecolor{sectcol}{rgb}{0.63,0.16,0.16}
\definecolor{palevioletred4}{rgb}{0.545,0.278,0.365}
\definecolor{gray40}{rgb}{0.40,0.40,0.40}
\definecolor{gray26}{rgb}{0.26,0.26,0.26}
\definecolor{olive}{rgb}{0.5,0.5,0.0}
\definecolor{gray78}{cmyk}{0,0,0,0.22}

\usepackage[
    pdftitle={Defining the role of open source software in research reproducibility},
    pdfauthor={Lorena Barba},
    pdfpagemode={UseOutlines},
    pdfpagelayout={TwoColumnRight},
    bookmarks, bookmarksopen,bookmarksnumbered={True},
    pdfstartview={FitH},
    breaklinks=true,
    colorlinks, linkcolor={linkcol},citecolor={linkcol},urlcolor={linkcol}
    ]{hyperref}

\makeatletter
\def\url@leostyle{%
  \@ifundefined{selectfont}{\def\UrlFont{\sf}}{\def\UrlFont{\footnotesize\sffamily}}}
\makeatother
\def\rrr#1\\{\par
\medskip\hbox{\vbox{\parindent=2em\hsize=6.12in
\hangindent=4em\hangafter=1#1}}}

\newlength{\up}
\newlength{\hup}
\begin{document}

\pagenumbering{arabic}
\renewcommand{\thepage} {\arabic{page}}

\thispagestyle{empty}

{ \Huge Defining the role of open source software in research reproducibility} 
\medskip

Lorena A. Barba, the George Washington University, Washington D.C. 

November 2021

\vspace{1cm}

\begin{abstract}
Reproducibility is inseparable from transparency, as sharing data, code and computational environment is a pre-requisite for being able to retrace the steps of producing the research results. 
Others have made the case that this artifact sharing should adopt appropriate licensing schemes that permit reuse, modification and redistribution. 
I make a new proposal for the role of open source software, stemming from the lessons it teaches about distributed collaboration and a commitment-based culture. 
Reviewing the defining features of open source software (licensing, development, communities), I look for explanation of its success from the perspectives of connectivism---a learning theory for the digital age---and the language-action framework of Winograd and Flores. 
I contend that reproducibility engenders trust, which we routinely build in community via conversations, and the practices of open source software help us to learn how to be more effective learning (discovering) together, contributing to the same goal.
\end{abstract}

\section*{What is reproducibility, and why does it matter?}
\vspace{\up}

Discussions in the scholarly community about reproducibility have intensified over the past decade, with a never-ending string of workshops, symposia, dedicated journal special issues like this one, and even Congressionally mandated studies conducted by the National Academies. 
The sister magazine of the IEEE Computer Society, \emph{Computing in Science and Engineering}, ran special issues on this theme in Jan/Feb 2009 and in Jul/Aug 2012. 
The guest editors of the latter highlighted that the extensive use of computation in the scientific endeavor has meaningfully changed the norms and standards of science-making. 
Computing is now an everyday activity for every scientist, in some form or another, from running simple statistics on empirical data, to running massive simulations on leadership computing facilities. 
The definitive, policy informing report of the \cite{nasem_2019} thus sharply focused its definition of \emph{reproducibility}: 
``obtaining consistent computational results using the same input data, computational steps, methods, and code, and conditions of analysis.'' 
This definition indelibly links reproducibility with transparency, via the requirement that data and code associated with a research study be made available to other researchers. 
It does not imply, however, that the code should be open source, just \emph{available}---this could possibly be under special permissions granted for the purpose of inspecting or reproducing the results. 
Yet, a large contingent of researchers who identify as members of the reproducible-research movement (including myself) vigorously advocate for open source software. 
In this article, I aim to elucidate the role of open source software in research reproducibility.

Why do we care about research reproducibility? 
The products of scientific knowledge are all around us, in the technology we use, how we protect from dangers, heal and live longer, entertain, communicate and travel, grow economies and build a better world. 
All this achievement is cumulative and hinges on a fabric of trust in the scientific process (we can see this clearly in cases of trust breakdown that have undermined science, e.g., the anti-vaccine movement and climate-change denialism). 
We use computer simulations, statistical analyses, and data-driven models to create scientific knowledge, and thus it is pertinent to ask when can we say that we have trustworthy evidence to justify the claimed findings. 
If we use simulations, for example, how do we show evidence that the results represent reliable data about the real world? 
How do we come to trust the computational results? 
It's interesting to note that questions like these were raised a few decades ago about \emph{experiments}. 
You may think it is rather obvious that we should believe the results obtained from an experiment that is directly observable, yet in many cases (e.g., in particle physics) experiments are far from indisputable. 
\cite{franklin1989} wrote about the neglect of experimental physics from the treatment of philosophers of science, and discussed the many strategies that experimental scientists use to provide grounds for rational belief in experimental results. 
For example: confidence in an instrument increases if we can use it to get results that are expected in a known situation. 
Or we gain confidence in an experimental result if it can be replicated with a different instrument or apparatus. 

In the direct detection of gravitational waves, arguably the greatest scientific breakthrough of our time, the LIGO collaboration spent months investigating every possible explanation for the signal captured by the apparatus before making their public announcement. 
If they told the world that the chirp was caused by two black holes merging a billion years ago, they had to be certain! 
To begin with, the collaboration used two facilities with the same apparatus, each costing hundreds of millions of dollars. 
The team's activities even included so-called blind injections: when artificial signals are sent to various groups of researchers to make them analyze the data and exercise the scientific workflow, arriving at the conclusion that the signals are not from gravitational waves. 
Every member of the collaboration is committed to the relentless quest for unimpeachable evidence. 
Notably, in their pursuance of trust, the collaboration founded the Gravitational Wave Open Science Center,\footnote{\url{https://www.gw-openscience.org/}} hosting data, software, documents, and web tools produced by the effort. 
The data and the multiple pieces of software are released under standard public licenses, usually Creative Commons Attribution (CC-BY4.0) for data and MIT or GPL licenses for code. 
And recently, \cite{brown2021} used the open data and code to reproduce the computational analysis that established the statistical significance of the detection. 
The authors also described several challenges they faced in the reproduction exercise, \emph{despite} the open availability of the original data and code. 
Among other hurdles, the original publication had not documented the version of the code used in the analysis, and the script used to make the final figure in the paper was not archived and shared. 
To finally succeed in reproducing the published result, the collaboration of one of the authors of the original paper, who had to resort to personal files, was needed. 
Even the most meticulous research team, with a broad commitment to open science, had missed something!

\section*{Open-source software in research}
\vspace{\up}

Researchers have been involved since the genesis of open-source software (OSS) about 50 years ago, and have been a driving force in OSS development ever since. 
Unix at Bell Labs, the Berkeley Software Distribution, the Unix-compatible GNU system at MIT, and Linux by Linus Torvalds at University of Helsinki are all examples of researcher-led precursors for today's OSS ecosystem.  
The term \emph{open source software} was introduced about 24 years ago \citep{peterson2008} and its definition put in black and white soon after by the Open Source Initiative (OSI).\footnote{\url{https://opensource.org}} 
It stipulates satisfaction of ten criteria, including not just availability of the source code but also the attachment of a license to use, modify and redistribute it freely. 
And herein lies one of the reigning misconceptions about open source software: simply sharing the source code in a hosting service, code repository or website is not enough---one must attach a standard public license approved by the OSI for the software to be called open source. 
Like with any creative work, copyright is automatically attached to computer code when it is written, and thus any source code posted online must be assumed ``All Rights Reserved'' by the copyright holder, unless it includes a license.

\begin{wrapfigure}{R}{0.5\textwidth}
\vspace{-0.6cm}
\begin{tcolorbox}[title=Copyleft vs. permissive OSS licenses, arc=0mm,,boxrule=0.5pt]
\footnotesize \textsf{
\emph{Copyleft} licenses require that any derivative works be under the same license of the original---also called ``share-alike.'' 
Some developers want to ensure open access to their work and all derivatives for  posterity. 
Although this may be considered virtuous in some circles, we should recognize that it is achieved by placing restrictions on the use of the software. 
(A typical copyleft license is GPL.) 
\emph{Permissive} licenses give more freedoms: the only restriction of use is usually that the original authors receive credit in any distribution of the software or any derivative works. 
Even commercial uses, or incorporating the software into other proprietary (closed) works, is allowed. 
Academic and research software benefits most from permissive licensing, enabling more impact and innovation. 
\cite{hunter2004} argued convincingly for the benefit of a permissive license, e.g., in encouraging industry users who would spurn share-alike code.
Typical examples are the BSD License, the MIT License and the Apache License.
In my research group, we use the BSD 3-clause license. It has a small difference in the wording about attribution compared to the MIT license, which uses the phrase ``substantial portions of the Software.'' Thus, a user could copy without attribution some of an MIT-licensed code (as long as the portion is deemed ``not substantial''). In research, we always prefer full attribution of any portions of reused works, and BSD 3-clause is more precise in this.}
\end{tcolorbox}
\end{wrapfigure}

Open-source licenses are an invention of great impact, as an alternative to intellectual-property restrictions. 
They allow people to coordinate their work freely, within the confines of copyright law, while making access and wide distribution their priority--this is fundamentally aligned with the norms of science, where we value academic freedom and wide dissemination of scientific findings. 
The key benefit of OSI-approved licenses is that researchers do not need extensive legal training or consultants to navigate these issues: 
the licenses are ``pre-packaged'' and ready to use. 
A researcher needs to know the conditions of their employment contract (whether the copyright belongs to the author or the employer), and be familiar with some basic license terms, e.g., as presented by \cite{morin2012}. 
With that information, they (or their employer) can choose one of the standard licenses for any particular software project. 
Sometimes the choice may be tinged with ideological positions, for example, in the alternative between a copyleft or permissive license. (See the sidebar.)

From the legal perspective, open source is a licensing model, as described above. 
When applied to research software, it contributes to transparency of the computational workflow and availability for others to use/modify/redistribute the software, one of the requisites of reproducibility. 
Transparency is a necessary but not sufficient condition, however: reproducible research implies the ability of another researcher or team to reuse the data, code and conditions of analysis of a published study. 
The key operating word here is \emph{reuse}, which as any researcher who has tried to exercise the digital artifacts of another will attest can be laborious and sometimes impossible. 
An open source license does not by itself support reuse; good-quality software design and documentation are needed in at least some measure. 
In support of software quality, the open source \emph{development model }takes center stage. 
Some guiding principles of the open development model are the mantra ``release early, release often,'' the idea of cultivating users as co-developers, and making full use of the Internet for distributed collaboration. 
Developing in the open is conducive to improved quality because users can make source-aware bug reports and cooperate with developers in more usefully identifying issues and devising solutions. 
When open source software projects nurture a community of users and contributors, problems can be identified and fixed more quickly, software design and quality improve over time, and user documentation can be made more friendly and complete. 
All of this helps make the software \emph{reusable}.
I should clarify that to satisfy reproducibility we may only need a weak form of software reuse: in the context of the original publication, to confirm the results. 
The stronger standard of software reuse that applies to libraries, for example, which have to work in many contexts, is not needed here. 
Even so, it is a bar seldom passed by research articles published today.

Some researchers remain apprehensive of the open development model, citing concerns that unskilled users may introduce unwanted changes or misuse the software leading to incorrect results. 
We should address another common misconception: developing in the open does not mean that others can make changes in your code willy-nilly. 
The communities of open source software have evolved coordinations, roles and tooling that expressly mediate distributed teams and contributors. 
Users outside of the original authors who are interested in the software make their own copy of the source code, and can make modifications in their copy. 
If they would like their modifications to be adopted by the original team, they make a move called a ``pull request,'' where they present their modified code to be peer reviewed and considered for merging into the original code base. 
If the original team has decided that they will consider such pull requests (a choice developers make ahead of time, depending on their staffing situation), team members taking the role of ``maintainers'' will review the proposed changes and engage in a conversation with the user-contributor to understand it, test it, request adjustments, and eventually approve or reject it. 
To facilitate this undertaking, developer communities have instigated formal processes and created enabling technologies. 
Teams increasingly adopt software testing methodologies, automate the processes of building from source and running the tests (called ``continuous integration''), and appoint leaders to conduct decision-making (i.e., adopt governance). 
These practices further contribute to maintaining and improving code quality and facilitating reuse of software. 
As for misuse of the code leading to wrong results, the scientific tradition of peer review has never quite eliminated erroneous results, but paired with transparency, the likelihood is that errors will be found sooner rather than later.

Another reason cited by researchers for working with closed, in-house code, and declining to release it even when submitting a scientific manuscript for review, is the time and effort they associate with this requirement. 
\cite{leveque2012} in their CiSE editorial cited a researcher survey: 78\% said they don't share code because of the extra work to clean up and document it. 
The question this raises is: can you trust the research results produced with such tangled and undocumented code? 
Since the times of the Enlightenment in the seventeenth century, the founders of empirical science stipulated that experiments should be recorded with painstaking detail: 
Francis Bacon and Robert Boyle ushered in a new era of carefully controlled and recorded experiments. 
The laboratory notebook has remained an indispensable companion of researchers for a few hundred years, not much changed until recent adoption of computational alternatives. 
``Good science requires good record keeping,'' according to the NIH guidelines \citep{nih2008guidelines}.
``Good records are complete, accurate and understandable to others. 
Records of research activities should be kept in sufficient detail to allow another scientist skilled in the art to repeat the work and obtain the same results.'' 
What many researchers balking at cleaning and documenting their research code fail to discern is that this failure could impact their own ability to retrace their steps, were their results challenged by a peer. 
Or it could impact the ability of a new team member to pick up where another (say, a graduate student) left off. 

The tools of open code development deliver built-in good record-keeping. 
In my experience, after using code version control for some time, one grows to see the value of applying it to everything: 
lab policies, web pages, documentation, personal notes, internal reports, article drafts, etc. 
The advantages are particularly apparent when more than one team member works on the same document or material: 
who added or deleted what, and when, is automatically captured. 
In a recent webinar for \emph{Nature} \citep{perkel2001}, I related our experience finding a small bug in the code that we used to generate all the results in a paper published recently. 
In fact, a collaborator's graduate student identified the bug in the process of learning about the method and following the code, thanks to it being openly available. 
After our initial alarm, it took just just a few days to download our own reproducibility packages from archival repositories (Zenodo and Figshare) and re-run all the computational experiments, to arrive at sufficiently close results that the main findings were unaltered (to our relief!). 
Having these detailed records and archives may have saved my team weeks or months of work trying to retrace the full workflow of the published work. 
Like lab safety and security measures that seem like a burden of extra labor until they are \emph{really} needed, reproducible research practices are insurance for that moment when a problem surprises us. 
I'm reminded of my favorite quote from \cite{donoho2009}: 
``… if everyone on a research team knows that everything they do is going to someday be published for reproducibility, they’ll behave differently from day one. Striving for reproducibility imposes a discipline that leads to better work.'' 

\section*{Science is a conversation}
\vspace{\up}

We tend to think of publication as the only medium for communicating scientific progress and findings, but science progresses also through preprint sharing, correspondence, conference interactions, social media, and any medium of conversation. 
Scientific knowledge is created in conversations among scientists, and using an expanded definition of conversation among scientists interacting with a body of knowledge (which is the product and record of other scientists' conversations). 
In this sense, science \emph{is} a conversation. 
Computational science places the computer as an agent in that conversation, and software is nothing more than the language (``code'') we use to interact with the machine and other researchers at the same time. 

In the last fifteen years or so, the framework of \emph{Connectivism} integrated theories of (human) learning with the new ways of interacting in the digital age. 
It defines connective knowledge as knowledge created and shared by an interacting community of individuals, or knowledge distributed across a network of connections. 
Under this lens, science is the formation of connections: between concepts, artifacts, people, theories, actions, etc. 
The role of the scientist is to be literate in this form of conversation, being able to traverse fluidly the network of connections and build the network further. 
\emph{Openness} is a property of networks that make them more successful for learning, that is, developing shared knowledge---other such properties are connectivity, diversity, and autonomy. 
Open sharing has a pedagogical purpose in this framework, as openness increases our capacity to create knowledge together. 
Quoting co-founder of Connectivism, Stephen Downes, talking about open education \citep{downes2017}: 
``Openness is about the possibilities of communicating with other people […] It’s not about \emph{stuff}, what you do with stuff. It’s about what you do with each other.'' 
Openness promotes rich networks, lively communities, and fertile connections. 
And this is good for science.

Open-source software projects, and the culture of their communities, have thus more to offer than a scheme for securing freedom from copyright restrictions on shared code. 
Open-source communities have developed templates for coordinating the actions of diverse groups of people, aimed at improving communication and working more effectively together. 
The ethos of open-source culture favors open development, networked collaboration, community around open-source projects, and a value-based framework. 
Gabriella Coleman \citep{coleman2012} researched open-source culture from an anthropological point of view, and observed that open-source projects build organizations with strong commitments to freedom of access, transparency, and joint governance and decision-making. 
In this setting is that open-source licenses were conceived, with the goal of enabling people to coordinate their work freely. 
With web technology playing a key role in those coordinations, platforms like GitHub gave form to structured conversations, while teams adopted rituals of collaboration and associated language, enabling them to create value together through productive interactions.

Thirty five years ago, Terry Winograd and Fernando Flores in their book \emph{Understanding Computers and Cognition} \citep{winograd-flores1986} talked about designing computer systems that can support humans to be more effective together. 
This was a groundbreaking textbook on system design, addressing how information technology could help improve human communication in organizations and in society. 
A key observation is that the coordinated action of people or teams happens in the context of \emph{conversations}. 
Language is not simply a vehicle for transmitting information about the world, but for changing the world: it has the dimension of acts and we invent the future via \emph{conversations for action}. 
This perspective helps explain the success of open-source collaboration practices. Take the ``pull request,'' for example. 
In this basic coordination of the open development model, a third-party (not belonging to the core team of developers) makes some changes to a code base and requests these changes to be merged into the official code repository. 
The changes could be a new feature or a bug fix. 
Notified of this request, members of the original software team review the changes, and may decide to accept the pull request, reject it, or ask for further changes in the third-party code. 
These moves are all standard in open-source communities, and everyone knows how to act depending on their role. 
Subject to time and effort constraints of those involved, a software project grows and improves thanks to contributions via pull request. 
GitHub structured the conversations around code that are pull requests, and gave tangible form to the premise of Winograd and Flores in this context. 
Thus, GitHub became a tool-of-the trade in the open-source world that supports the workflow, and promotes a culture of collaboration. 

The open-source standard workflows have other examples of the commitment-based culture of collaboration, mediated by web platforms such as GitHub. 
Consider the ``issue tracker'': a project's contribution policy may ask to ``log an issue for any question or problem.'' 
Implied in this policy is a commitment by the maintainers to coordinate actions for addressing the issue, even if it is to mark it as postponed for later review. 
In fact, a public version control repository is a \emph{signal} that the project is open to users participating in the development process, either simply watching the progress, or filing bug reports in the issue tracker, or offering code changes via pull request. 
It is common to view a public issue tracker as a strong indicator of the project's commitment to openness. 
For this reason, having an open issue tracker is a requirement for a submission to The Journal of Open Source Software (http://joss.theoj.org/) to proceed to peer review, for example. 
From the perspective of Winograd and Flores, and the conversations-for-action model, when in open-source projects we talk about building community, we are also talking about building commitment. 
And by this we mean the promises that get things done: people making promises to each other that take care of their concerns, changing the direction of the future. 
Quoting Fernando Flores: ``When people coordinate successfully in [a conversation for action] … action happens, trust is enhanced, and relationships are strengthened.''

\section*{Reproducibility as a trust-building endeavor}
\vspace{\up}

Reproducibility leaders Jeff Leek and Roger Peng wrote: ``To maintain the integrity of science research and the public’s trust in science, the scientific community must ensure reproducibility and replicability by engaging in a more preventative approach that greatly expands data analysis education and routinely uses software tools.'' \citep{leek-peng2015} 
Both scientific integrity and general trust in science are often linked with reproducibility. 
At the same time, a crisis narrative often percolates through discussions during this period of escalating concern with reproducibility and replicability (the past twelve years or so). 
Why? In part because the broad discussion was catalyzed by highly public fiascos, such as the failure to replicate research on genomic signatures of lung cancer, leading to the uncovering of misconduct by Anil Potti at Duke University.
It is in these instances where trust breakdowns occur that we begin paying more sharp attention to trust-building activities. 
Add in the rapid transition to computation and data becoming central to the scientific enterprise, and we have a credibility crisis, according to \cite{donoho2009} (and many others). 
Inaction would amount to ``blind trust'' or denial (self-deception), which is corrosive to integrity. 
So we, the computational science community, have been debating what changes we need to make to our practice to increase trustworthiness. 
We have thus focused on the traits of studies that can be called reproducible: they are published transparently, they share data and code, and they adopt good practices of data management and record-keeping. 
Next, we find ourselves debating what to do with the shared digital artifacts of research: should they be peer reviewed, alongside to a research manuscript? 
Should journals exercise the computational workflow, and confirm that the results can be reproduced? 
Soon, we are facing monumental hurdles associated to the effort and cost of such implementation of reproducible research (especially in the context of large-scale or high-performance computing).

In such deliberations, it is worthwhile to reflect on what it is that we're trying to achieve when demanding that published research be reproducible. 
As previously stated, at the core is transparency: researchers should always provide all needed information for another to be able to reproduce their results. 
The research not only has to be published, but also \emph{conducted} in such a way that reproducibility is realizable. 
It is uncertain to what extent the scientific community needs to exercise the research artifacts, in the review process or post peer review, to confirm the reproduction of results. 
As discussed in the \cite{nasem_2019} report (see Finding 4--5), assessments of the state of reproducibility of published works usually fall back on assessing transparency, as a proxy. 
The committee assessed that ``determining the extent of issues related to computational reproducibility across fields or within fields of science and engineering is a massive undertaking.''
Similarly, directly reproducing all results presented for peer review would involve a  gargantuan effort. 
What is the point, then, of making all digital artifacts—primary and secondary data, analysis code, and metadata—available and open? 
I propose that in essence reproducibility is an endeavor aiming to build trust. 
Scientific findings are supported by evidence, but one of the shared values of the scientific ethos is ``organized skepticism''---this is one of the four so-called Mertonian norms; see \cite{christensenETal2019}.
We aspire for the scholarly record to be trustworthy, and thus allocate credibility building activities, even if in principle we might say that the ``facts speak for themselves.'' 
Arguably, peer review was historically introduced to similarly increase trust in the published literature.

But trustworthiness and trust are not the same thing. 
\cite{solomon-flores2003} [p.76 ff.] explain that trustworthiness is a trait that can be demonstrated, established with reasons and evidence. 
Trust, the act of trusting, hangs on the experience and frame of mind of the one who trusts, the other party in the equation. 
Trust is dynamic and social: it is cultivated through commitment and action, through conversation. 
For this reason, I do not believe it is possible to develop IT products that fully automate reproducibility, and offer a ``one-click solution.'' 
We cannot ``solve'' the reproducibility problem with any approach that takes away the responsibility of the human participants. 
In this argumentation, I am influenced by the  deliberate definitions of trust by \cite{solomon-flores2003}. 
The precise meaning of trust is affected by context: whether it be interpersonal relationships, business, or politics, the meaning specializes to that setting. 
Here we are discussing the conduct of science, and our collective trust in the findings or results (not on the researcher, personally) and ultimately trust in the scientific institution. 
Regardless of context, however, we can distinguish between \emph{simple} trust, and \emph{authentic} trust. 
We often think of simple trust first, upon hearing the word: that basic, unthinking trust that is taken for granted, trust by default, absence of suspicion, without scrutiny or reflection. 
This kind of trust is a poetic illusion, and it rarely exists. 
``Authentic trust is both reflective and honest with itself and others. […] Authentic trust is not opposed to distrust so much as it is in a continuing dialectic with it'' \citep[p.92]{solomon-flores2003}. 
Thus, it is not a paradox to say that scientific facts are objectively true, and to also say that to trust research findings we expect to rely on a network of information shared and acts performed. 
It is perhaps more like a \emph{polarity}: two seeming opposites that need each other. 
Authentic trust is \emph{rational}, deriving from both actions and commitments. 
Importantly, trust is a way of dealing with complexity in an ever more complex world, say \cite{solomon-flores2003} [p.9]. 
Scientific research has also dramatically increased in complexity. 
Even in social sciences \citep{christensenETal2019}, empirical research can involve thousands of lines of code. 
Since currently none of this code ends up in the published research article, many in the scientific community are calling for research code to be made publicly available alongside the paper. 
Yet, simply dumping megabytes of spaghetti code on GitHub is not helpful. 

In a talk titled ``Science as Amateur Software Development'' \citep{mcelreath2020} anthropologist and evolutionary ecologist Richard McElreath links trust to open source software as follows.
He spoke from his own experience as a researcher and also a member of open source software communities.
At the 49:22 time mark,  he draws an analogy between software engineering methods of unit testing and continuous integration, and empirical science workflows. 
From expressing a theory as a probabilistic program, using an algorithm to prove that the analysis will be able to identify causal effects, and testing the pipeline with synthetic data sets, and doing all this with standard open source methods, ``now you’re ready and we trust your pipeline; it’s time to put real data in it [and] of course it’s important that all of this history be open and available in a public repository so that people trust the analysis.''
He later says:  ``the big problem [...]\ in common between the endeavor of science and the endeavor of developing open source software to support science is in integrating work from different experts and doing it in a responsible way, and doing it transparently, in public so that people who come after us can can have some trust in what we've done and in our work and also when mistakes are discovered [...]\ they can go back and find the source of the mistake and correct it and and learn from that...''

Most of us who participate in communities of open source software for science come to the same realization: we are learning to work better together. 
At some stage, you start having reservations about papers you read not only because they are published without accompanying code and data, but because the code shows no signs of being developed in the open model. 
On the other hand, for large-scale projects involving computations that cannot be reproduced due to cost or availability of the computing facility, we are more comfortable trusting the results if the history of the project displays open development practices since inception.
Open source software development helps manage complex projects with high quality, injecting professionalism in addition to promoting transparency. 
It thus gives reproducibility a boost.

\section*{Concluding remarks}
\vspace{\up}

Reproducibility is essentially an activity that builds trust, making possible a more effective interdependency of research results and associated artifacts. 
Adopting open source software in science gives rise to a form of social contract and way of working that builds trust. 
Research is impactful when the network of results and how they are shared have a meaningful role in society and/or the continuing accumulation of knowledge, with researchers participating productively. 
When computational science is part of this activity, how does using open source software affect that? 
By itself, it may not affect much at all, but if it is the model for ``how we do things,'' how we work in collaborations large and small, then it can have a meaningful role---looking at it with a connectivist lens. 
The model for how we work with research software becomes: when we are writing and running code, we feel a responsibility, we are committed to ensure this knowledge is injected into the network, to be interacted with. 
In this model, we see software and data as the artifacts we create to learn with, and to learn with others. 
This has an effect on the researchers who are writing code and running it to produce results, growing their sense of value in the science enterprise. 
It reflects the change in culture we need for valuing software and data, and publishing reproducible research.


{\small
\bibliography{./oss_in_rr}
\bibliographystyle{jponew}
}

\end{document}